\documentclass{article}
\usepackage{amsmath,amssymb,graphicx,natbib}

\begin{document}

\title{On the relationship between ODEs and DBNs}
\author{C.J.Oates\,$^{1,2,3}$$^*$, S.M.Hill\,$^{1,2}$ and S.Mukherjee$^{3,2,1}$\footnote{to whom correspondence should be addressed}
\\
$^{1}$Centre for Complexity Science, University of Warwick, CV4 7AL, UK \\ 
$^{2}$Department of Statistics, University of Warwick, CV4 7AL, UK \\
$^{3}$Netherlands Cancer Institute, 1066 CX, Amsterdam, The Netherlands.}

\maketitle

Recently, Li {\it et al.} ({\it Bioinformatics} {\bf 27}(19), 2686-91, 2011) proposed a  method, called Differential Equation-based Local Dynamic Bayesian Network (DELDBN), for reverse engineering gene regulatory networks from time-course data. 
We commend the authors for an interesting paper that draws attention to the close relationship between 
dynamic Bayesian networks (DBNs) and differential equations (DEs).
Their central claim is that modifying a DBN to model Euler approximations to the gradient rather than expression levels themselves is beneficial for network inference. The empirical evidence provided is based on time-course data with equally-spaced observations. However, as we discuss below, in the particular case of equally-spaced observations, 
Euler approximations and conventional DBNs
lead to equivalent statistical models that, absent artefacts due to the estimation procedure, yield networks with identical inter-gene edge sets.
Here, we discuss further the relationship between DEs and conventional DBNs and present new empirical results on unequally spaced data which demonstrate that modelling Euler approximations in a DBN can lead to improved network reconstruction.

A dynamic Bayesian network (DBN) is a Bayesian network (i.e. a graphical model based on a directed acyclic graph) with an explicit time index. 
Consider biochemical time-series expression data $X_i(t)$ where $i \in \{ 1 \ldots p \}$ indexes genes (or other molecular variables of interest) and $t \in \{ 1 \ldots T \}$ indexes time. 
In the DBNs considered in \cite{Li} (with linear Gaussian conditionals and edges only from one time slice to the next)
mean expression at time $t+1$ is modelled as a linear function of expression at time $t$
\begin{eqnarray}
X_i(t+1) \sim N\left( \sum_{j=1}^p \tilde{\beta}_{ij} X_j(t) , \sigma^2_i \right)
\label{eq: linear}
\end{eqnarray}
where $\tilde{\beta}_{ij}$ is a parameter describing the influence of gene $j$ on target $i$,  $\sigma_i$ is a noise parameter and $N(\mu,v)$ denotes a Gaussian distribution with mean $\mu$ and variance $v$. For convenience we refer to models of the form in Eqn. \ref{eq: linear} as ``conventional'' DBNs.

In contrast, consider a DE that models the gradient as a deterministic linear function with parameters $\beta_{ij}$:
\begin{eqnarray}
\frac{dX_i(t)}{dt} = \sum_{j=1}^p \beta_{ij} X_j(t)
\label{eq: ODE}
\end{eqnarray}
DELDBN combines DBNs (Eqn. \ref{eq: linear}) with DEs (Eqn. \ref{eq: ODE}) by using an Euler gradient approximation
\begin{eqnarray}
\frac{X_i(t+1)-X_i(t)}{\Delta t} \sim N\left( \sum_{j=1}^p \beta_{ij} X_j(t),\sigma^2_i\right)
\label{eq: difference}
\end{eqnarray}
where $\Delta t$ denotes the interval, in units of time, between observations with time indices $t$ and $t+1$.

In the terminology of regression, the left hand sides of Eqns. \ref{eq: linear} and \ref{eq: difference}  are responses that are modelled using predictors $X_j$, with independent samples indexed by $t$.
The main claim of Li {\it et al.} is that improved performance in network reconstruction may be achieved by modelling the response as the Euler gradient (Eqn. \ref{eq: difference}) rather than the observed value (Eqn. \ref{eq: linear}) of gene expression at a given time, provided the time interval $\Delta t$ between samples is not too large. 
Evidential support for this conclusion is provided using data from a synthetic gene regulatory network ``IRMA'' that was constructed in {\it Saccharomyces cerevisiae} \citep{Cantone}. 

DELDBN carries out inference regarding network topology  using 
Markov blankets, 
facilitated by a heuristic search implemented in the R package \texttt{BNLearn} \citep{Scutari}. The Markov blanket for a node $i$ in a Bayesian network is the set of nodes comprising  $i$'s parents, its children, and its children's other parents. For the DBNs considered in Li {\it et al.} the Bayesian network is bipartite and a target $i$ has no children, only parents (as depicted in Fig. 1 in Li {\it et al.}). Therefore the Markov blanket $\mathrm{MB}(i)$ is identified with the parents of $i$ in the DBN. According to the Euler gradient model (Eqn. \ref{eq: difference}) the Markov blanket of a target $X_i$ is given  by the non-zero coefficients
\begin{eqnarray}
\mathrm{MB}_{\text{Euler}}(X_i) = \{ X_j : \beta_{ij} \neq 0 \}.
\label{eq: MB difference}
\end{eqnarray}

When the time $\Delta t$ between samples is constant, the Euler gradient model is simply a reparameterisation of the conventional DBN:
\begin{eqnarray}
X_i(t+1) = X_i(t) + \Delta t \sum_{j=1}^p \beta_{ij} X_j(t) = \sum_{j=1}^p \tilde{\beta}_{ij} X_j(t) \\
\tilde{\beta}_{ij} = \left\{ 
     \begin{array}{lr}
       \Delta t \beta_{ij} & i \neq j \\
       1 + \Delta t \beta_{ij} &  i = j
     \end{array}
   \right.
\label{eq: tilde}
\end{eqnarray}
In this case of equally spaced observations, under the conventional DBN the Markov blanket is
\begin{eqnarray}
\mathrm{MB}_{\text{C-DBN}}(X_i) = \{ X_j : \tilde{\beta}_{ij} \neq 0 \}
\end{eqnarray}
which differs from Eqn. \ref{eq: MB difference} only in the possible presence/absence of a self-loop $X_i$, since for $i \neq j$ we have that $\beta_{ij} \neq 0$ if and only if $\tilde{\beta}_{ij} \neq 0$. Thus, for equally spaced observations the Markov blanket is invariant (up to inclusion of a self-loop) to the reparameterisation that occurs in going from a conventional DBN to Euler gradient responses. This means that the two formulations are equivalent with respect to inference regarding the inter-gene edge set when equal sampling intervals are used. Therefore in order to distinguish between the reverse-engineering performance of these approaches, it is essential to consider the regime in which data are sampled unevenly in time. However, results reported by Li {\it et al.} were obtained using only data sampled evenly in time\footnote{It is interesting to ask why the output of DELDBN  on the IRMA data appeared to improve using Euler gradient responses. The growth-shrink (GS) algorithm \citep{Margaritis} was used to infer Markov blankets from data. GS proceeds by carrying out conditional independence tests, based in this case on the Pearson correlation coefficient $r$ and statistic
$r \sqrt{\frac{T-2}{1-r^2}} \sim \mathcal{T}_{T-2}$,
where $\mathcal{T}_{T-2}$ denotes the $t$-distribution with $T-2$ degrees of freedom. This particular approximate approach to Markov blanket identification while computationally efficient is not invariant to the reparametrisation relating the conventional DBN and Euler gradient models.
Since these models are structurally equivalent with respect to inter-gene edges, this suggests that the improved performance of DELDBN on inter-gene edges reported in Li {\it et al.} may be an artefact due to the specific estimator used.}.

Nevertheless, we fully agree that the relationship between dynamics and DBNs merits careful investigation. In particular many time-course datasets in bioinformatics are obtained with unequal  sampling intervals. Then, the equivalence between conventional DBNs and Euler gradient models does not hold, making the choice of formulation an important question. 
We therefore undertook empirical comparison of a range of modelling approaches,
which may all be viewed as variations of the well studied variable-selection problem in linear regression \citep{Oates}. 
This subsumes both conventional DBNs and the Euler gradient model discussed above.

\begin{figure}[h]
\centering
\includegraphics[trim = 0.75cm 8cm 1cm 8cm,clip,scale = 0.4]{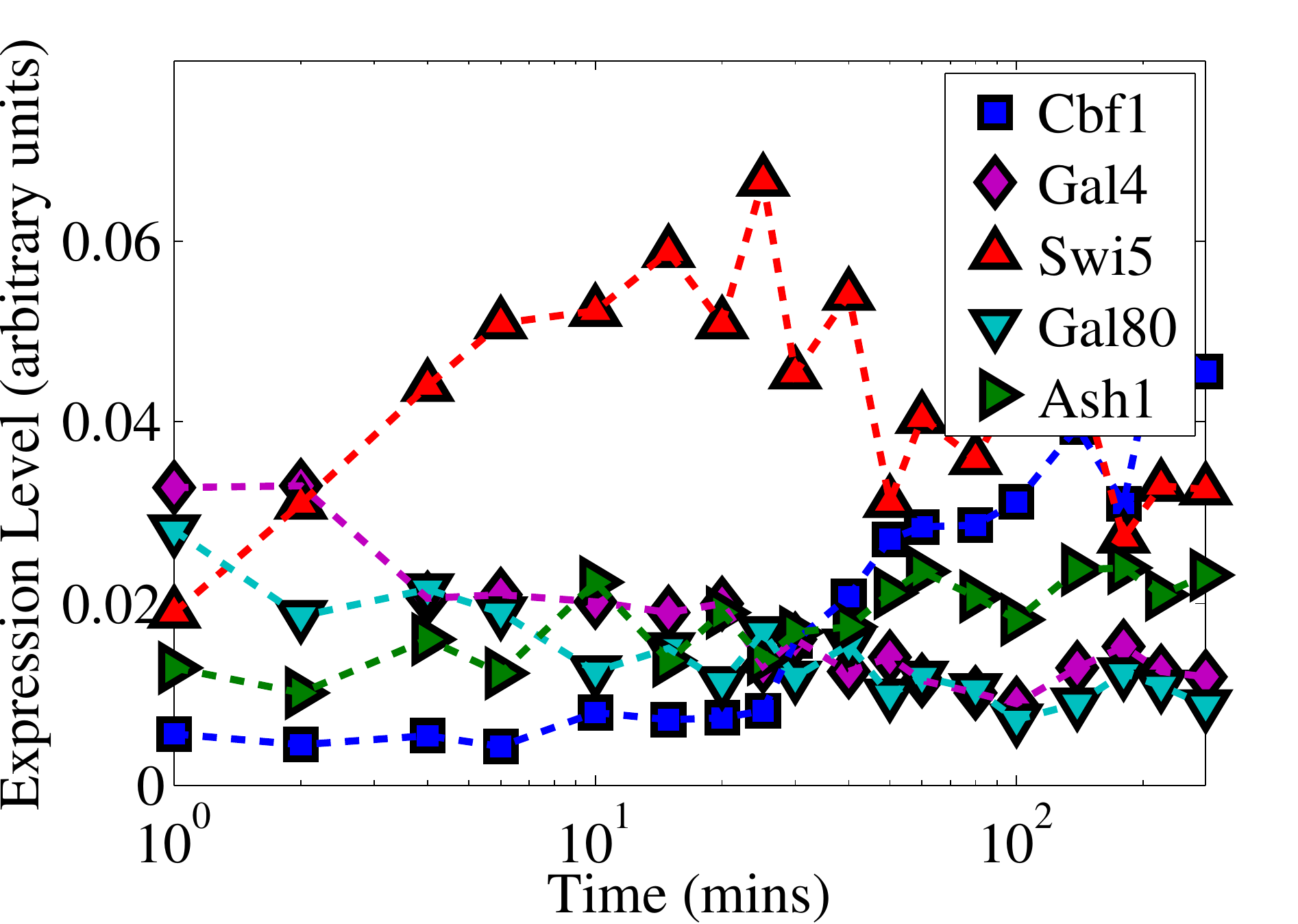}
\caption{Simulated data, typical dataset. The DE model of \cite{Cantone} was used to simulate data from the IRMA network at uneven time intervals.}
\label{fig: data}
\end{figure}

In order to obtain unevenly sampled data from the  IRMA network studied in Li {\it et al.}, we used the differential equation (DE) model described by \cite{Cantone}. This model has been demonstrated to provide good fit to the IRMA data. We generated data at unevenly spaced times 0, 1, 2, 4, 6, 10, 15, 20, 25, 30, 40, 50, 60, 80, 100, 140, 180, 220 and 280 minutes, adding Gaussian measurement error with variance set to give a signal-to-noise ratio equal to 20. A typical dataset generated in this way is presented in Figure \ref{fig: data}.

We carried out inference within a  Bayesian framework. 
Denote by $y$ the responses (for target $i$, for simplicity we suppress dependence on $i$ in what follows), either as in a conventional DBN or Euler approximations. Let $D$ denote the design matrix constructed according to the Markov blanket $\mathrm{MB}$ (for notational simplicity we leave dependence on $\mathrm{MB}$, i.e. the graph structure, implicit below), and let $m$ be the number of columns in $D$. Assuming additive Gaussian error we get
$y = D \beta + \epsilon, \; \epsilon \sim N(0,\sigma^2I)$
where $I$ denotes the identity matrix,  $\beta$ collects together all coefficients and $\sigma^2$ is the variance of the error.  
We score models by integrating the corresponding likelihood against a prior for $(\beta,\sigma^2)$.
Here, we used a g-prior \citep{Zellner} 
$\beta \mid \sigma^2 \sim N(0,\sigma^2 T (D'D)^{-1})$ for coefficients and 
$\pi(\sigma^2) \propto 1/\sigma^2$,
where $n$ is the sample size in the regression sense. 
This leads to closed-form marginal likelihood
\begin{eqnarray}
\pi(y \mid \mathrm{MB}) & \propto & \left(\frac{1}
{1+T}\right)^{m/2} \left[ y'y - 
\left(\frac{T}{1+T}\right) \hat{y}'\hat{y} 
\right]^{-(T-1)/2}
\end{eqnarray}
where $\hat{y} = D(D'D)^{-1}D' y$. 
Network inference is carried out by Bayesian model averaging, 
using the posterior probability 
\begin{eqnarray}
\mathbb{P}(j\text{ regulates }i) =  \sum_{\mathrm{MB}} \frac{ \mathbb{I}\left\{X_j \in \mathrm{MB}(i)\right\} \pi(y \mid \mathrm{MB}) \pi(\mathrm{MB})}{\sum_{\mathrm{MB}'} \pi(y \mid\mathrm{MB}') \pi(\mathrm{MB}')}  
\end{eqnarray}
to score  a directed edge from gene $j$ 
to  target $i$, where $\mathbb{I}\{A\}=1$ is $A$ is true, otherwise $\mathbb{I}\{A\}=0$.

In experiments below we take a network prior which, for each target 
$i$, is uniform over the number of predictors $m_i$ up to a maximum 
permissible in-degree $d_{\max}$, that is 
$\pi(\mathrm{MB})  \propto  \prod_i  {p \choose m_i}^{-1} \mathbb{I}\left\{m_i \leq d_{\max}\right\}$, 
but note that richer network priors are available in the literature \citep{Mukherjee}.
A Markov blanket estimator is obtained by thresholding 
posterior edge probabilities; for threshold $\tau$ this gives a network with estimated edge-set
$\hat{E} =  \{ (j,i) : \mathbb{P}(j\text{ regulates }i) \geq \tau \}$.
For small maximum in-degree $d_{\max}$, exact inference by enumeration 
of variable subsets may be possible. Otherwise, Markov chain Monte 
Carlo (MCMC) methods can be used to estimate posterior edge 
probabilities \citep{Ellis,FriedmanIII}. In the experiments 
here we use exact inference by enumeration, with $d_{\max} = 2$.

\begin{figure}[h]
\centering
\includegraphics[trim = 2cm 6.5cm 2cm 6.5cm,clip,scale = 0.4]{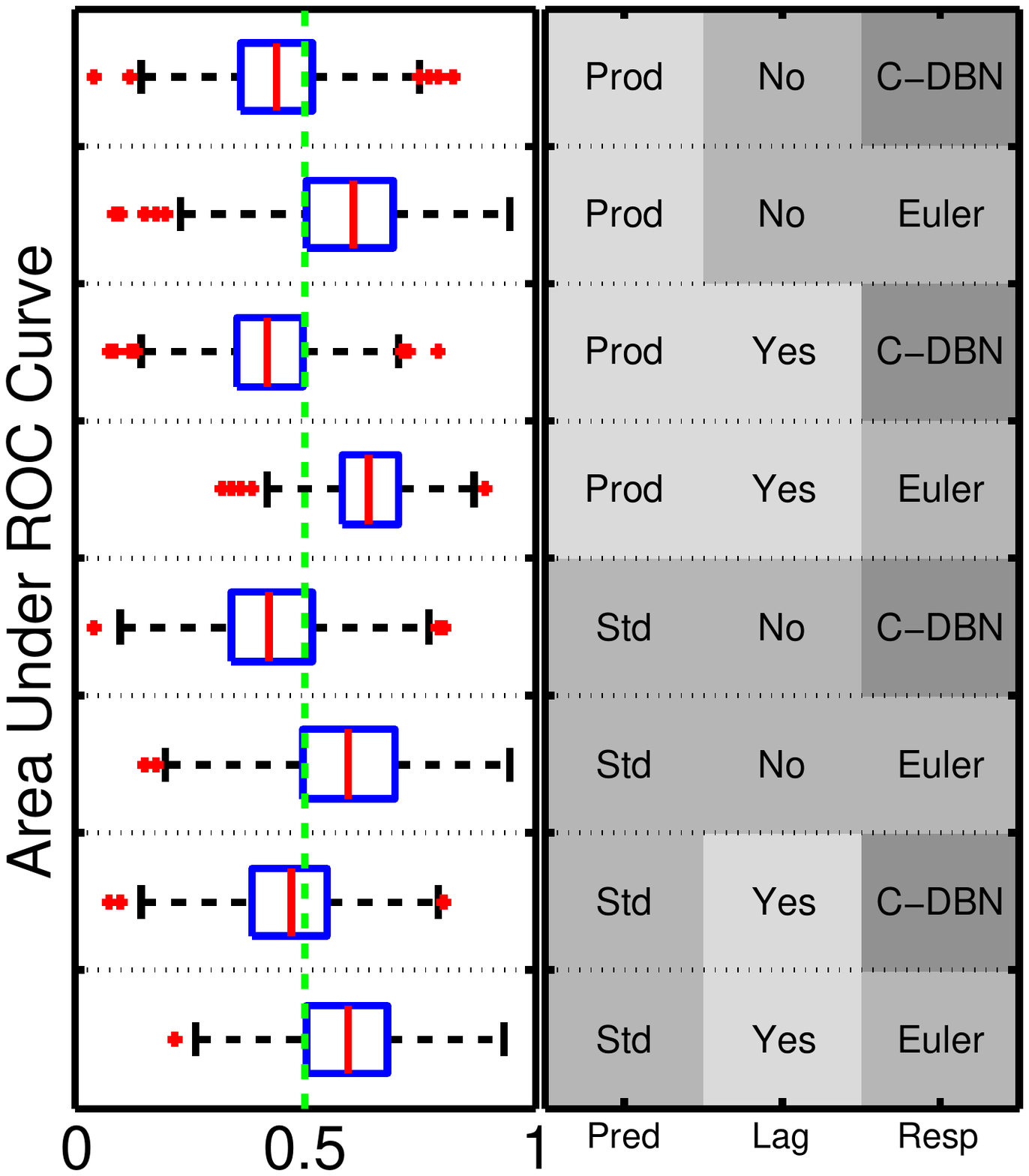}
\caption{IRMA network, area under receiver operating characteristic curves (AURs). AURs were calculated for a range of modeling approaches, based on data simulated from the IRMA network  at uneven time intervals. Key: ``Resp"  - [Euler] Euler derivative approximations / [C-DBN] conventional DBN; ``Pred" -  [Prod] using products $X_iX_j$ of predictors / [Std] otherwise;  ``Lag" -  [Yes] additional lagged predictors / [No] otherwise.}
\label{fig: AUR}
\end{figure}

Under the framework outlined above, we assessed network reconstruction using both conventional DBNs and Euler gradients. Since the DE model of \cite{Cantone} is nonlinear, it is natural to also investigate whether the use of products $X_jX_k$ of predictors (in addition to linear predictors) improves network reconstruction, by capturing some nontrivial aspects of the dynamics. Similarly, as suggested in the discussion of Li {\it et al}, since the DE model of \cite{Cantone} contains a delay term, it is interesting to investigate whether the use of lagged predictor variables  improves performance. The approaches we considered can be summarised as:
\begin{center}
		\begin{tabular}{l|r}
		    Predictor set & \{ Standard, Product \} \\
			Lagged predictors & \{ No, Yes (lag $\approx$ $T/10$) \} \\
			Response & \{ Conventional DBN, Euler gradient\}\\
		\end{tabular}
\end{center}
Performance was assessed using the area under receiver operating characteristic curves (AUR), equivalent to the probability that a randomly selected true edge has a higher score than a  randomly selected false edge; higher values of AUR correspond to better performance.
We carried out inference for  1000 sampled datasets for each method to obtain distributions over AUR scores, as shown in Figure \ref{fig: AUR}.
Inference based on the Euler gradient response outperforms the conventional DBN, supporting the central claim of Li {\it et al.}
The use of products of predictors together with the inclusion of lagged predictors led to slightly improved performance.

In summary, we presented empirical evidence that Euler approximations to dynamics coupled with DBNs can be useful in reverse engineering gene regulatory networks. Furthermore we showed how such models may be viewed in a regression framework, for which there exists a wide literature on variable selection. However our investigation was somewhat idealised, since in practice data are often obtained under destructive sampling and averaging over large numbers of cells. An extended discussion on the relationship between cellular dynamics, nontrivial observation processes and linear regression may be found in \cite{Oates}.

\section{Funding} We gratefully acknowledge support from EPSRC EP/E501311/1 (CJO, SMH \& SM) and NCI U54 CA 112970-07 (SM).

\end{document}